\documentclass[usenatbib]{aa}  

\usepackage{graphicx}
\usepackage{txfonts}
\usepackage{siunitx}
\usepackage{soul}
\usepackage{longtable}
\usepackage{booktabs}
\usepackage[dvipsnames]{xcolor}

\usepackage[colorlinks,citecolor=blue,urlcolor=blue,filecolor=blue,linkcolor=blue]{hyperref}
\begin{document}

   \title{The impact of Solar wind variability on pulsar timing}
   \titlerunning{The impact of Solar wind variability on pulsar timing}

   \author{C. Tiburzi\inst{1}\thanks{tiburzi@astron.nl}, G. M. Shaifullah\inst{1,2}, C. G. Bassa\inst{1}, P. Zucca\inst{1}, J.\ P.\ W.\  Verbiest\inst{3,4}, N. K. Porayko\inst{4}, E. van der Wateren\inst{1,5}, R. A. Fallows\inst{1}, R. A. Main\inst{4}, G. H. Janssen\inst{1,5}, J. M. Anderson\inst{6,7}, A-.S. Bak Nielsen\inst{4,3}, J. Y. Donner\inst{4,3}, E. F. Keane\inst{8} J. K\"unsem\"oller\inst{3}, S. Os\l{}owski\inst{9,10}, J-.M. Grie\ss meier\inst{11,12}, M. Serylak\inst{13,14}, M. Br\"uggen\inst{15}, B. Ciardi\inst{16},  R.-J. Dettmar\inst{17}, M. Hoeft\inst{18}, M. Kramer\inst{4,19}, G. Mann\inst{20}, C. Vocks\inst{20}
                    }
\authorrunning{C. Tiburzi et al.}

\institute{
	ASTRON $-$ the Netherlands Institute for Radio Astronomy, Oude Hoogeveensedijk 4, 7991 PD Dwingeloo, The Netherlands
	\and Dipartimento di Fisica ``G. Occhialini'', Universit\`a di Milano-Bicocca, Piazza della Scienza 3, I-20126, Milano, Italy
	\and Fakult\"at f\"ur Physik, Universit\"at Bielefeld, Postfach 100131, 33501 Bielefeld, Germany
	\and Max-Planck-Institut f\"ur Radioastronomie, Auf dem H\"ugel 69, 53121 Bonn, Germany
	\and Department of Astrophysics/IMAPP, Radboud University, P.O. Box 9010, 6500 GL Nijmegen, The Netherlands
    \and Technische Universit\"at Berlin, Institut f\"ur Geod\"asie und Geoinformationstechnik, Fakult\"at VI, Sekr. H 12, Stra\ss e des 17. Juni 135, 10623 Berlin, Germany
    \and GFZ German Research Centre for Geosciences, Telegrafenberg, 14473 Potsdam, Germany
    \and SKA Organisation, Jodrell Bank, Macclesfield SK11 9FT, UK
    \and Gravitational Wave Data Centre, Swinburne University of Technology, P.O. Box 218, Hawthorn, VIC 3122, Australia
    \and Centre for Astrophysics and Supercomputing, Swinburne University of Technology, P.O. Box 218, Hawthorn, VIC 3122, Australia
     \and LPC2E - Universit\'e d’Orl\'eans / CNRS, 45071 Orl\'eans cedex 2, France
     \and Station de Radioastronomie de Nan\c{c}ay, Observatoire de Paris, PSL Research University, CNRS, Univ. Orl\'eans, OSUC,18330 Nan\c{c}ay, France
    \and South African Radio Astronomy Observatory, 2 Fir Street, Black River Park, Observatory 7925, South Africa
    \and Department of Physics and Astronomy, University of the Western Cape, Bellville, Cape Town 7535, South Africa
    \and Hamburger Sternwarte, University of Hamburg, Gojenbergsweg 112, 21029 Hamburg, Germany
    \and Max-Planck-Institut f\"ur Astrophysik, Karl-Schwarzschild-Stra\ss e 1, 85748 Garching b. M\"unchen, Germany
    \and Ruhr University Bochum, Faculty of Physics and Astronomy, Astronomical Institute, 44780 Bochum, Germany
    \and Th\"uringer Landessternwarte, Sternwarte 5, 07778 Tautenburg, Germany
    \and Jodrell Bank Centre for Astrophysics, University of Manchester, M13 9PL, UK
    \and Leibniz-Institut f\"ur Astrophysik Potsdam (AIP), An der Sternwarte 16, 14482 Potsdam, Germany
    }

   \date{Received MM DD, YYYY; accepted MM DD, YYYY}
 
  \abstract
   {High-precision pulsar timing requires accurate corrections for dispersive delays of radio waves, parametrized by the dispersion measure (DM), particularly if these delays are variable in time. In a previous paper we studied the Solar-wind (SW) models used in pulsar timing to mitigate the excess of DM annually induced by the SW, and found these to be insufficient for high-precision pulsar timing. Here we analyze additional pulsar datasets to further investigate which aspects of the SW models currently used in pulsar timing can be readily improved, and at what levels of timing precision SW mitigation is possible.}
   {Our goals are to verify: a) whether the data are better described by a spherical model of the SW with a time-variable amplitude rather than a time-invariant one as suggested in literature, b) whether a temporal trend of such a model's amplitudes can be detected.}
   {We use the pulsar-timing technique on low-frequency pulsar observations to estimate the DM 
   and quantify how this value changes as the Earth moves around the Sun. Specifically, we monitor the DM in weekly to monthly observations of 14 pulsars taken with parts of the LOw-Frequency ARray (LOFAR) across time spans of up to 6 years. We develop an informed algorithm to separate the interstellar variations in DM from those caused by the SW and demonstrate the functionality of this algorithm with extensive simulations. Assuming a spherically symmetric model for the SW density, we derive the amplitude of this model for each year of observations.
}
   {We show that a spherical model with time-variable amplitude models the observations better than a spherical model with constant amplitude, but that both approaches leave significant SW induced delays uncorrected in a number of pulsars in the sample. The amplitude of the spherical model is found to be variable in time, as opposed to what has been previously suggested.}
   {}

   \keywords{pulsars:general, solar wind, ISM: general, gravitational waves}

   \maketitle
%

\section{Introduction}\label{Sec:intro}

High-precision pulsar timing \citep{lk04} is a technique used to, for example, investigate irregularities in the Solar system planetary ephemerides (e.g. \citealt{cgl18,vts20}), generate alternative time-scale references (e.g. \citealt{hgc20}), test general relativity (e.g. \citealt{agh18,vcf20}) and alternative theories of gravity (e.g. \citealt{sck13}) or search for low-frequency gravitational waves with Pulsar Timing Arrays (PTAs, e.g. \citealt{tib18,bst19}). In particular, a level of timing residuals below 100\,ns is usually indicated as the white-noise threshold to achieve in PTA experiments (e.g., \citealt{jhm15,sej13}).

The sensitivity of these experiments can be significantly degraded by various noise processes such as those caused by errors in clock standards or inaccuracies in the planetary ephemerides.

One of the most common sources of noise in pulsar-timing data \citep{lsc16} is the variable amount of free electrons along the line of sight (LoS). The radio waves coming from pulsars are dispersed due to the ionized medium, leading to time delays depending on the observing frequency, following the relation:
\begin{equation}\label{eq:deltat}
    \Delta t = \frac{e^2}{2 \pi m_e c}\:\frac{DM}{f^2} = \mathcal{D}\:\frac{DM}{f^2}
\end{equation}
where $\Delta t$ is the time-delay (in s) induced at an observing frequency $f$ (in MHz) with respect to infinite frequency, $e$ is the electron charge, $m_e$ is the electron mass and $c$ is the speed of light ($e^2/2 \pi m_e c$ is the \textit{dispersion constant} $\mathcal{D} = 1/(2.4\times10^{-4})~\mathrm{MHz^2}~\mathrm{pc}^{-1}~\mathrm{cm}^3~\mathrm{s}$, see \citealt{mt72}) and DM is the dispersion measure (in pc cm$^{-3}$):
\begin{equation}
    DM = \int_{LoS} n_e(l)\:\mathrm{d}l
\end{equation}
with $n_e$ being the electron density along the LoS. 

Variations in $n_e$ along the LoS induce DM fluctuations, causing changing $\Delta t$ contributions to the arrival times of pulsar radiation that need to be taken into account in pulsar timing experiments. The two main contributions to the DM along a certain LoS (for a comprehensive review, see \citealt{lcc16}) are the ionized interstellar medium (IISM) and the Solar wind (SW). In particular, the SW contribution to DM depends on the Solar elongation of the pulsar (i.e., the projected angular separation between the pulsar and the Sun), whose temporal variations therefore induce DM time-fluctuations.

The SW has been recognized as a noise source that could induce false detections of gravitational waves in PTA experiments \citep{thk16}. 

The standard pulsar timing approach to mitigate the SW contribution (e.g., the International PTA data releases by \citealt{vlh16} and \citealt{pdd19}) typically consists in approximating the SW as a spherically symmetric distribution of electrons $n_{e, sw}$ \citep{ehm06}:
\begin{equation}\label{eq:swdensitymod}
    n_{e, sw} = \mathrm{A_{AU}}\left [ \frac{1\mathrm{AU}}{r}\right]^2
\end{equation}
where $r$ is the distance between the pulsar and the Sun and $\mathrm{A_{AU}}$ is the free electron density of the Solar wind at 1~AU, reported to be $7.9$ and time-constant \citep{mca19}, and which we will henceforth refer to as the \textit{amplitude} of the SW density model. The DM contribution of this model is obtained by integrating Equation~\ref{eq:swdensitymod} along the LoS, and can be expressed as \citep{ehm06,yhc07}:
\begin{equation}\label{eq:dmsunsph}
    DM_{sw} = 4.85 \times 10^{-6} \mathrm{A_{AU}} \frac{\rho}{\sin\rho}~\mathrm{pc~cm^{-3}}
\end{equation}
where $\rho$ is the pulsar-Sun-observer angle. This model implicitly assumes that the amplitude $\mathrm{A_{AU}}$ is constant with time and independent of the ecliptic latitude of the pulsar. To account for the fact that this model may not be an optimal SW approximation at small Solar elongations, it is common to eliminate data points taken at close ($<5$~degrees) angular distances from the Sun in pulsar timing experiments \citep{vlh16}.  

However, the SW is more complex than implied from this simple model. Under Solar minimum conditions it is mostly bi-modal, with a \textit{fast} stream seen above polar coronal holes, and a \textit{slow} stream seen above a mostly-equatorial streamer belt \citep[e.g.][]{col96}. A polar coronal hole can sometimes extend towards equatorial latitudes, allowing the fast and slow streams to interact, leading to denser regions of compression at the leading edge of the fast stream and rarefied regions following behind \citep[e.g][]{schwenn90}. Coronal Mass Ejections (CMEs) will further complicate this picture. The picture becomes even more complex as Solar activity increases towards a maximum, when the bi-modal structure effectively breaks down allowing coronal streamers to extend to high latitudes. The reader is referred to \citet{schw06b} for a more detailed illustration of the SW system.

The shortcomings of the spherical model were neatly illustrated in a recent publication by \citet{ttt20} who observed the Crab pulsar in 2018 using the Toyokawa Observatory. This paper also compared their results to SW conditions assessed from observations of interplanetary scintillation and coronal white light, and provided a useful discussion on how observations of pulsars could also be used to assist research into the SW.

The aforementioned shortcomings led \citet{yhc07a} to propose a revised SW model for the pulsar DM which considered the SW as bimodal. The authors used different free-electron radial distributions for each of the two SW streams, and used Solar magnetograms to decompose the LoS into parts affected by one or the other component, the total contribution from the Solar-wind being the sum of these individual contributions.  This was demonstrated in the paper to better correct the DM for the SW contribution than the basic spherical model, but it should be noted that the pulsar observations were taken during the approach to solar minimum, when a bimodal solar wind structure is more evident.

\citet{tvs19} compared the performance of these models on highly-sensitive, low-frequency observations of PSR~J0034$-$0534, while also allowing a time-variable amplitude (following the approach of \citealt{ych12}) in both of the models. The authors demonstrated that neither model provided an adequate description of the SW impact on the dataset, but also that the spherical one performed better than the other.  The observations used in that paper were, however, taken at solar maximum, which may explain why the bimodal model did not perform better in that instance. More explanation of the possible reasons is given in \citet{tvs19}. 

In this article we expand on the analysis of \citet{tvs19}, using a larger sample of pulsars to verify a) whether the SW DM contributions to these pulsar LoSs are better described by a spherical model with a time-variable amplitude or a time-constant one as suggested in the literature \citep{ehm06,mca19}, and b) whether a consistent temporal trend can be detected in the amplitudes of this model which might suggest that the bimodal approach be revisited in future work.

The article is structured as follows: in Section~\ref{Sec:dataset} we present the dataset and the selected pulsars, while in Section~\ref{Sec:dataanalysis} we describe the analysis. The results are presented in Section~\ref{Sec:resdis}. In Section~\ref{Sec:future} we discuss future prospects, and in Section~\ref{Sec:conclusions} we draw our conclusions.
\section{Dataset}\label{Sec:dataset}

The utilized dataset comes from a number of pulsar monitoring campaigns carried out with the high-band antennas of different subsets of the International LOFAR (LOw Frequency ARray) telescope \citep{vwg13,sha11}: the six German International LOFAR stations, the Swedish International LOFAR station, and the LOFAR Core. The observing bandwidth covers a frequency interval from ${\sim}100$ to ${\sim}190$~MHz, with a central frequency of about $150$~MHz (variations of a few MHz occur among the different observing sites). The recorded data were coherently dedispersed, folded into 10-second-long subintegrations modulo the pulse period, and divided into frequency channels of 195~kHz with the \textsc{DSPSR} software suite \citep{vb11}. The integration length ranges from $1$ to $3$ hours with the international stations, and from $7$ to $20$ minutes with the LOFAR core (for more details regarding the observational setup, see \citealt{pnt19,dvt19} and \citealt{tvs19}).

\begin{figure*}
    \centering
    \includegraphics[scale=0.85,trim={3cm 1cm 3cm 0.5cm}]{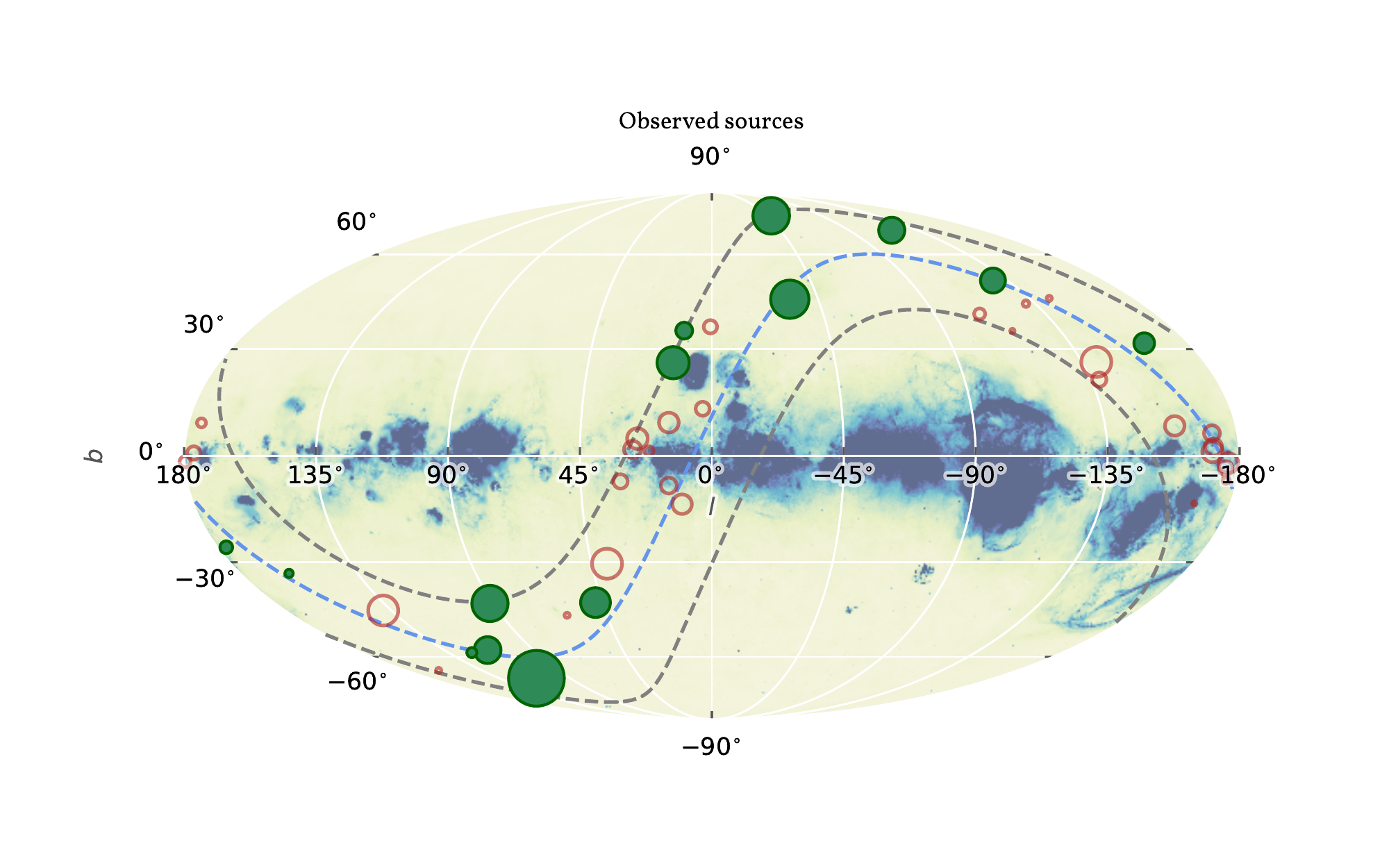}
    \caption{Sky distribution in Galactic coordinates of selected (filled green circles) and rejected (unfilled red circles) pulsars from the lists in Table~\ref{Tab:kept} and Table~\ref{Tab:sources_nosw}. Larger markers denote higher (median) precision of the measured DM. The dashed blue line marks the ecliptic and the dashed gray lines show a region of $\pm$20~degrees from the ecliptic. The background shows the merged all-sky H$\alpha$ map from \citet{fin03}.}
    \label{Fig:skyplot}
\end{figure*}

Together, the aforementioned observing campaigns monitor more than $100$ pulsars. However, for the scope of this article, we selected pulsars with the following characteristics: a) ecliptic latitude between $-20$ and $+20$~degrees, b) observing cadence higher than once per month, c) more than one year of observing time-span, d) without gaps between successive observations exceeding $100$ consecutive days. This results in a dataset of $43$ pulsars, whose sky locations are shown in Figure~\ref{Fig:skyplot} and characteristics are reported in tables~\ref{Tab:kept} and \ref{Tab:sources_nosw}\footnote{Part of the reported values come from the ATNF pulsar catalog, \url{https://www.atnf.csiro.au/research/pulsar/psrcat/} \citep{mht05}.}, and for which we have used all the data available until May $2019$ (up to August $2019$ in some cases). This source list was further refined to 14 sources that prove useful probes of the Solar wind, as discussed in Section~\ref{Sec:final}.

\section{Data analysis}\label{Sec:dataanalysis}
In the following, we detail the analysis methods applied to the data to obtain the DM values, disentangle the IISM-induced effects, and identify the pulsars showing persistent SW signatures. We make use of the pulsar timing technique, which keeps track of every pulsar rotation to model the evolution of the pulse period and phase over the timespan of our dataset. This allows us to average each observation in time and improve the signal-to-noise ratio (S/N). 

\subsection{Calculation of the DM values}\label{Sec:dmestimates}
For all observations, independent of the observing site, we removed radio-frequency interference, corrected for azimuth and elevation-dependent gain using the LOFAR beam model within the \textsc{dreambeam} package\footnote{\url{https://github.com/2baOrNot2ba/dreamBeam}}, and applied band-limitations to retain a common frequency range between ${\sim}118$ and ${\sim}188$~MHz\footnote{Band-limiting is necessary to avoid biases in our results because the observing sites record slightly different original bandwidths. The indicated frequency range is common to all of the observing sites (for more details, see \citealt{dvt19,tvs19}).}. These operations were carried out using the \textsc{psrchive} software suite \citep{hvm04,vdo12} and a modified version\footnote{\url{https://github.com/larskuenkel/iterative_cleaner}, see also \citet{kue17}.} of the \textsc{Coastguard} software suite \citep{lkg16}.

For each pulsar we then computed a DM value per observation through the pulsar timing technique by proceeding as follows. We first selected the dataset that covered the longest time-baseline. These observations were weighted by the square of their S/N, added together and then fully averaged in time and partially in frequency (usually down to $10$ frequency channels to increase the S/N\footnote{For particularly faint pulsars we applied a larger frequency-averaging factor.}). Finally, the template was smoothed using a wavelet smoothing scheme first introduced by \citet{dfg+13}. We then collected the observations obtained by all the observing sites for that pulsar, we fully averaged them in time and partially in frequency to the same resolution of the template. For each observation we then generated a set of ToAs associated to the frequency channels with the \textsc{psrchive} software suite by cross-correlating the observation with the reference template.

A few exceptions to the aforementioned general template-generation scheme were adopted in case the pulsar was too faint, or bright but strongly affected by red noise (due to irregularities in the pulsar rotation or extreme IISM-linked DM variations, that caused the template to appear broadened). In the first case, we averaged the longest dataset described earlier over frequency and time to then produce an analytic template, obtained by approximating the data-derived pulse profile with a sum of von Mises functions. In the second case, we used a small subset of phase-aligned observations which were subsequently averaged and smoothed. 

After generating a set of frequency-resolved ToAs per observation, we used the \textsc{tempo2} software suite for pulsar timing \citep{ehm06} and the procedure outlined in \citet{tvs19} to calculate one DM value per observation\footnote{Note that, as for \citet{tvs19}, the DM time derivatives included in the original timing model for that pulsar were only applied to properly dedisperse the template and frequency-average the observations, but they were not used in the subsequent determination of the DM variations.}. We then combined the DM time series from all the available observing sites, after subtracting the reference DM value.

\subsection{Disentangling the IISM contribution}\label{Sec:disentangl}
The obtained DM time series show variations due to both the SW and the IISM. To remove the influence of the IISM, for each pulsar we proceeded as follows. 

The DM time series was divided into 460-day long segments centered on the Solar conjunctions, i.e., adding an additional 1.5 months of baseline to a 6-month time window on either side of the Solar conjunction. Hence, the segments overlap for about 100~days. Segments that either contained gaps of more than 55 days between successive observations, or those where the effective time span is less than 368 days (80\% of 460 days) were also discarded, as they do not provide a long enough baseline to properly define the IISM effects (this only affects initial and final segments).

The DM time series in each segment was modeled in a Bayesian framework as the sum of a spherically-symmetric SW model (Equation~\ref{eq:dmsunsph}) with $\rm{A_{AU}}$ being a free parameter, and a polynomial to account for the IISM contribution. By simulating DM time-series affected by Kolmogorov turbulence \citep{ars95}, we found that a cubic polynomial was sufficient to model the DM variations due to the IISM on 460-day long segments (see the method's validation in Appendix~\ref{Asimulations}). To account for any common systematic error in the estimation of the DM uncertainties, we inserted an additional parameter in the model, summed in quadrature with the DM uncertainties in the likelihood function. We used a Markov Chain Monte Carlo method \citep[implemented using the {\it emcee} package;][]{fhlg13} to obtain the parameters of the cubic polynomial, the uncertainties correction parameter and the SW amplitude $A_\mathrm{AU}$, and account for the covariance of the SW amplitude with the parameters of the IISM model. We assigned flat priors to the polynomial coefficients, and a flat and positive prior to the amplitude of the SW model and the uncertainties correction parameter. 

As a final step, the modeled IISM contribution was subtracted from the DM time series. In the overlapping region between two successive years, data points that lay within 6 months of the preceding Solar approach were approximated by the plasma model computed for the first year, while data points within 6 months of the following Solar approach were approximated by the one computed for the second year\footnote{The overlap between adjacent years guarantees a continuous and smooth IISM model across multiple years.}. After estimating the model, data points collected earlier than 6 months before the first Solar conjunction, and later than 6 months after the last Solar conjunction were discarded. 

\begin{figure*}
\centering
\includegraphics[scale=0.35,trim={0 0 0 0}]{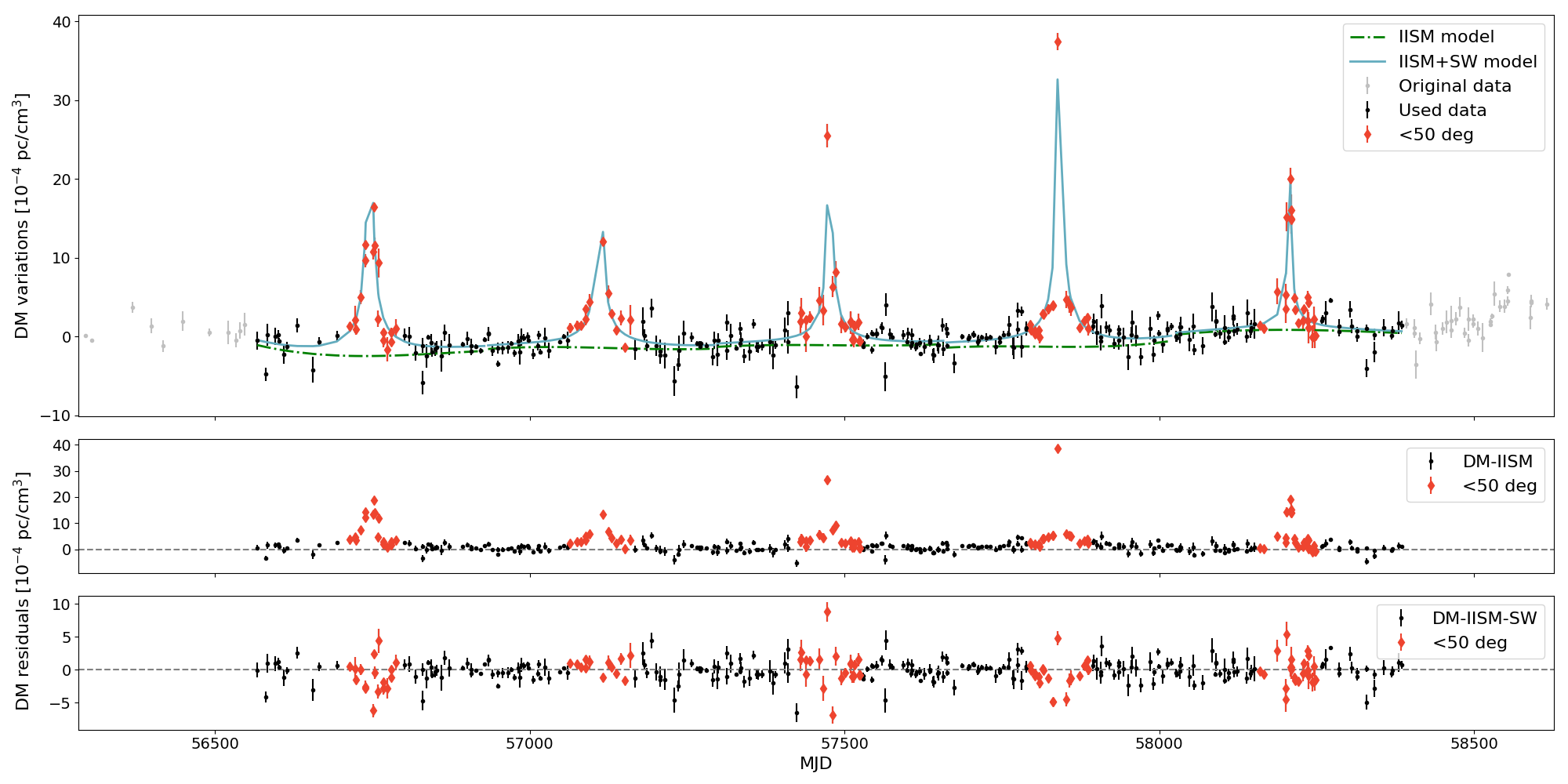}
\caption{IISM-SW disentanglement in PSR~J0030+0451. The upper panel shows the time series of the DM variations and the IISM models (dot-dashed green line) and of the SW and IISM combined (solid blue line). The used data are represented as black dots and red diamonds (where the red diamonds indicate the observations within 50 degrees in Solar elongation and the black dots the observations beyond 50 degrees), while the discarded data points are in gray. The middle panel shows the DM residuals after subtraction of the IISM model only, and the lowest panel shows the DM residuals after subtraction of both the IISM and the SW models.}\label{Fig:disentanglement}
\end{figure*}

As an example of the final result, Figure~\ref{Fig:disentanglement} shows the IISM disentanglement for PSR~J0030+0451.

We note here that in \citet{tvs19} the authors demonstrated that the spherical model is a poor description of the SW, when tested against sufficiently sensitive data. Nevertheless, we have adopted it in the procedure described above for two reasons. First, it was proven to be better among the two available models compared in that work. Secondly, in \citet{tvs19} we found that, while the spherical approximation could not model the short-term SW-induced DM variations, it provided a reasonable description of the long-term ones.

\subsection{Final selection}\label{Sec:final}
The procedure outlined above allowed us to refine our pulsar selection by rejecting those sources where the SW signature is not reliably detected. Specifically, we retained a pulsar if and only if in more than half of the dataset: a) the model described in the previous section was preferred over an IISM-only model as evaluated by the Bayesian information criterion (BIC), b) the posterior distribution of the SW model's amplitude was significantly different from zero\footnote{In PSR~J1300$+$1240 only the first two years (out of a total  five) meet the requirements. However, we included it in the final selection after visually inspecting the DM time series and manually examining the results.}.

\begin{figure*}
\centering
\includegraphics[scale=0.5,trim={2.1cm 0 0 0}]{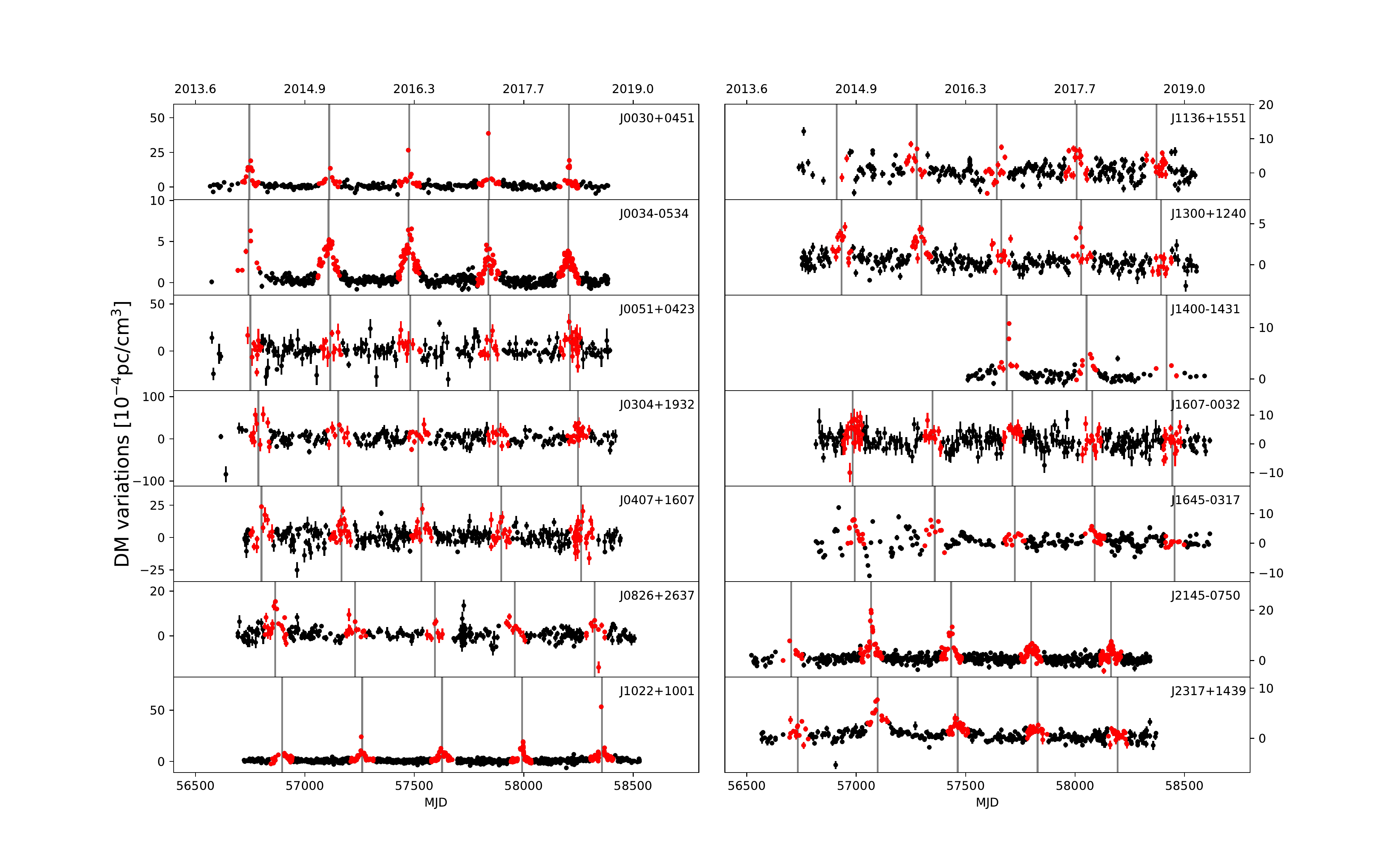}
\caption{DM time series of the selected pulsars, after subtraction of the IISM component. Observations highlighted in red were taken when the Solar elongation of the source was less than $50$~degrees, while the gray vertical lines mark the MJDs of the Solar conjunctions. For the sake of visual clarity, these plots show only 95\% of the most precise measurements although all data points are used in the analysis.}\label{Fig:sources}
\end{figure*}

\begin{figure}
    \centering
       \includegraphics[scale=0.55,trim={0cm 0cm 0cm 0cm}]{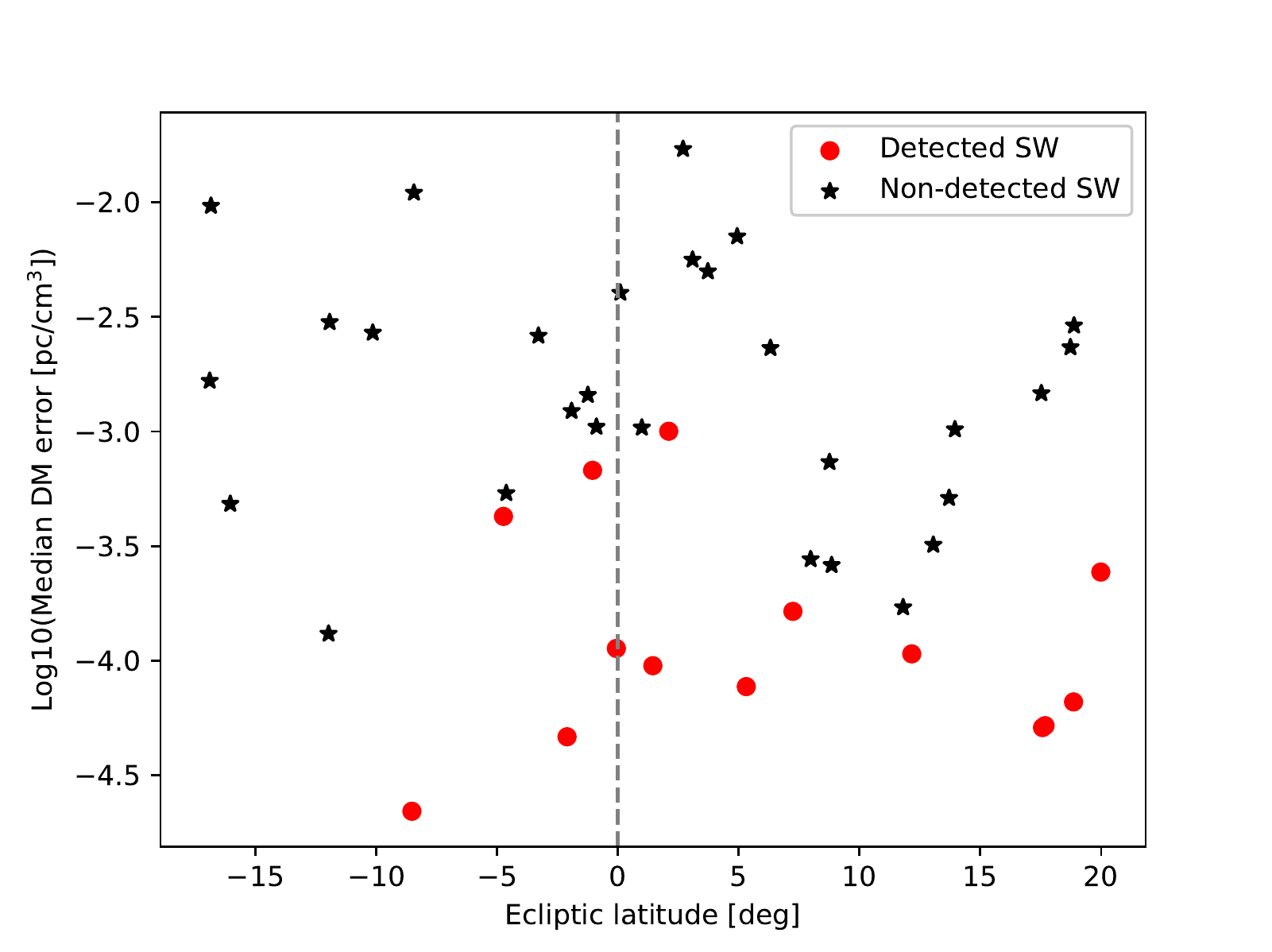}
    \caption{Pulsars included in the final selection (red circles) versus the excluded pulsars (black stars) as a function of ecliptic latitude and the logarithm of the median DM error.}
    \label{Fig:precisionelat}
\end{figure}

A total of $14$ pulsars satisfied the mentioned requirements, as reported in Table~\ref{Tab:kept} (for the discarded sources, see Table~\ref{Tab:sources_nosw} in the Appendix). Among these there are the PTA-class millisecond pulsars J0030+0451, J0034$-$0534, J1022+1001, J2145$-$0750 and J2317+1439 \citep{bdd19}, which, as expected, display the best DM precision of the sample. We consider the 14 selected sources as well-suited for studying the electron density in the SW at low frequencies in the Northern hemisphere. Figure~\ref{Fig:precisionelat} shows that, for equal ecliptic latitude, the pulsars included in the final selection always present the best DM precision among the sources at the same Ecliptic latitudes. The few cases of sources with high DM precision where the SW is not detected, can be explained typically by a combination of low S/N and a poor sampling cadence (e.g., PSR~J1024$-$0719). For the few pulsars in which these causes are not applicable (e.g., PSR~J0837+0610), we speculate that the reason lies in an asymmetry of the Solar wind contribution with respect to the heliographic latitude. 
A more rigorous test of this hypothesis will be presented in a future work through comparisons with data-derived magneto-hydrodynamic simulations of the SW electron density fluctuations, such as EUHFORIA \citep{pp17}.

Figure~\ref{Fig:sources} shows the SW component of the DM variations of the final selection.

\begin{table*}
\centering
\caption{Final source selection, encompassing $14$ pulsars. The table reports the source name, the covered time-span, the number of observing sites that have been monitoring that specific source, the Galactic coordinates, the rotational spin period, ecliptic latitude, dispersion measure (DM) of the pulsar, as measured during the general pulsar timing analysis described in Section~\ref{Sec:dataanalysis}, the decimal logarithm of the median DM uncertainty and the number of observations used to generate the data-derived template.}\label{Tab:kept}
{\small
\begin{tabular}{cccS[table-format=2.1]S[table-format=2.1]S[table-format=4.1]S[table-format=2.2]S[table-format=2.1]S[table-format=2.2]S[table-format=3]}
\hline
Name & Time-span & Observing & \multicolumn{2}{c}{Galactic} &\multicolumn{1}{c}{Period} & \multicolumn{1}{c}{Ecliptic} & \multicolumn{1}{c}{DM} & \multicolumn{1}{c}{Log} & \multicolumn{1}{c}{Observations}\\
&           & Sites & \multicolumn{2}{c}{Coo. [deg]} &\multicolumn{1}{c}{[ms]} & \multicolumn{1}{c}{Latitude [deg]} & \multicolumn{1}{c}{[pc/cm$^3$]} & \multicolumn{1}{c}{M(eDM)} & \multicolumn{1}{c}{per template} \\
\hline
J0030+0451		&	2013-01 \quad 2019-05 & 7 & 113.1& -57.6 & 4.9	& 1.45  & 4.3   & -4.02 &215  \\
J0034$-$0534 		&	2012-12	\quad 2019-05 & 6	& 111.5& -68.1 & 1.9 	& -8.53 & 13.8  & -4.66 &227  \\ 
J0051+0423 		  &	2013-08	\quad 2019-05&   5& 	 123.&	-58.5	&	354.7 & -1.05 	& 13.9   & -3.17  & 208 \\ 
J0304+1932   &	2013-08	\quad 2019-05&   6&	 161.1&-33.3	&	1387.6 & 2.10 	& 15.7   & -3.0   & 208        \\ 
J0407+1607   		&	2013-08	\quad 2019-05 & 7	& 176.6& -25.7 & 25.7 	& -4.74 & 35.6  & -3.37 &216  \\ 
J0826+2637   &	2013-08	\quad 2019-05 & 7 	& 197.0& 31.7  & 530.7 	& 7.24 	& 19.5  & -3.79 &150  \\ 
J1022+1001   		&	2012-12	\quad 2019-08 & 8	& 231.8& 51.1  & 16.5 	& -0.06 & 10.3  & -3.95 &290  \\ 
J1136+1551   &	2013-08	\quad 2019-05 & 6 	& 241.9& 69.2  & 1187.9 & 12.16 & 4.8   & -3.97 &299  \\ 
J1300+1240   &	2012-12	\quad 2019-08 & 6 & 311.3& 75.4  & 6.2 	& 17.58 & 10.2  & -4.29 &239  \\ 
J1400$-$1431 		&	2015-10	\quad 2019-05 & 7 & 327.0& 45.1  & 3.1 	& -2.11 & 4.9   & -4.33 &167  \\
J1607$-$0032 &	2013-09	\quad 2019-05&   6& 10.7&	35.5	&	421.8 & 19.99 	& 10.7   & -3.61  & 26         \\
J1645$-$0317 &	2013-08	\quad 2019-08 & 7 & 14.1& 26.1   & 387.7 	& 18.86 & 35.8  & -4.18 &4    \\
J2145$-$0750 		&	2013-01	\quad 2019-08 & 7 & 47.8& -42.1  & 16.1 	& 5.31 	& 9.0   & -4.11 &232  \\
J2317+1439   		&	2012-12	\quad 2019-08 & 8 & 91.4& -42.4  & 3.4 	& 17.68 & 21.9  & -4.28 &7    \\
\hline 
\end{tabular}
}
\end{table*}

\section{Results}\label{Sec:resdis}
\subsection{Performance with respect to a SW model with constant amplitude}
\begin{figure*}
    \centering
    \includegraphics[scale=0.7,trim={0 0 0 0}]{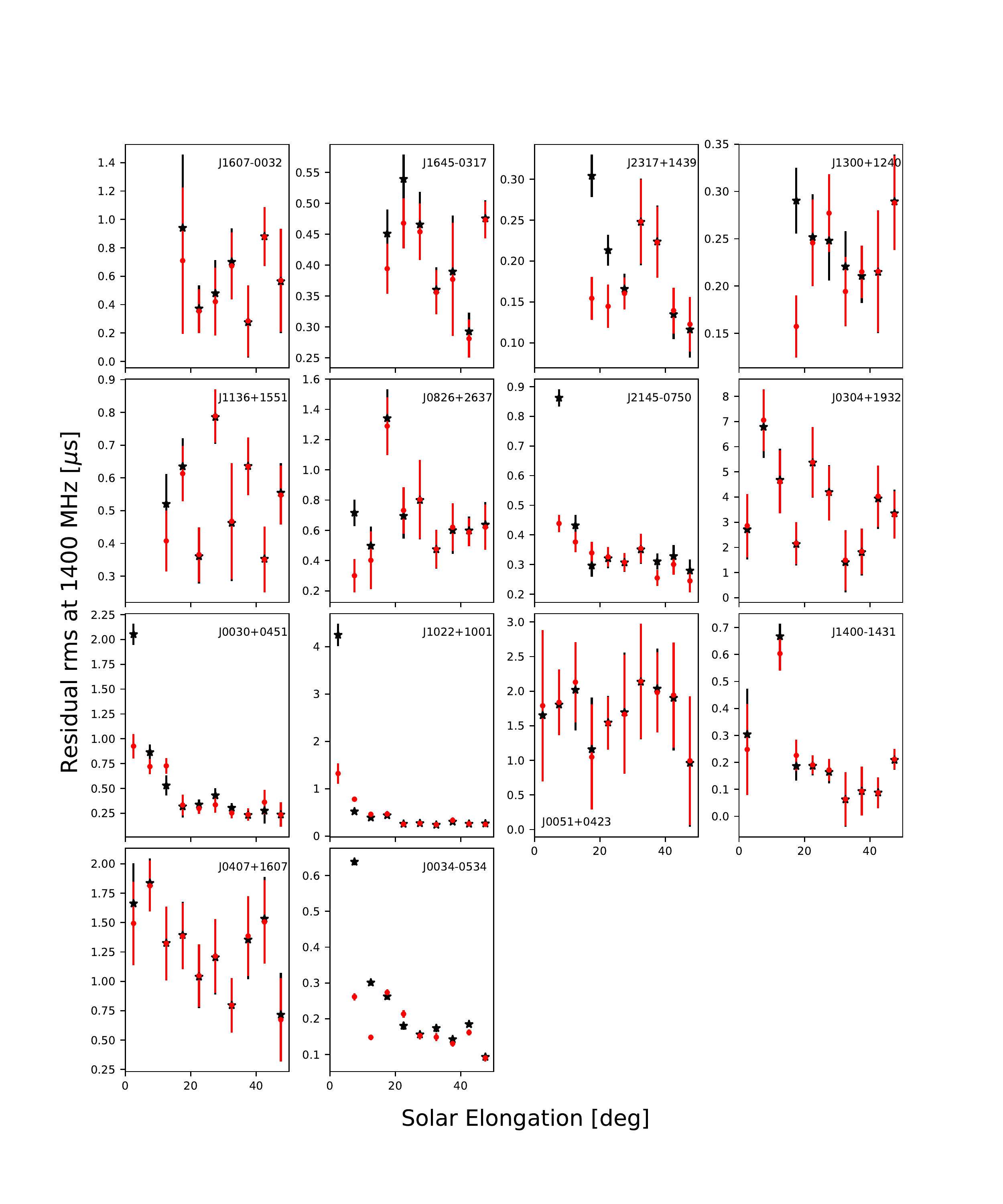}
    \caption{Root-mean-square of the residual timing delays in our data, rescaled to an observing frequency of 1400~MHz. The timing residuals were binned in solar elongation with a resolution of 5 degrees and a Solar-wind model with time-constant (black) or variable (red) amplitude was subtracted. The pulsars are sorted by ecliptic latitude.
    }
    \label{Fig:modelcomparison}
\end{figure*}

From the point of view of pulsar timing experiments, it is important to understand whether a spherical model of the SW performs better (i.e., yields smaller residuals when subtracted from the observations) when a time-variable or a static amplitude is assumed. For this aim, we repeated the analysis described in Section~\ref{Sec:disentangl} on the final pulsar selection by fixing the amplitude for the spherical SW model to a value of $7.9~\mathrm{cm}^{-3}$ \citep{mca19} and we compared their results. We stress that, for this analysis, we exclude those segments in the pulsar's datasets where the posterior distribution of the SW amplitude was found to be consistent with zero in the previous Section. Assuming an absence of frequency-dependence of the DM (cf. \citealt{css16}), in Figure~\ref{Fig:modelcomparison} we show the comparison between the two spherical models, with a time-dependent and a time-invariable amplitude, reported through Equation~\ref{eq:deltat} in terms of residual time delays at 1400~MHz (the main reference frequency for high-precision pulsar-timing studies). In particular, we display the rms of the time delays induced by the residual DM fluctuations in the two different analyses, binned in Solar elongation. In Figure~\ref{Fig:modelcomparison}, the black dots and red stars refer respectively to a constant- and a variable-amplitude SW model.

By assuming the rms of the residual time delays as criterion, a model with a variable amplitude performs better than one with a constant amplitude in more than 60\% of the cases for Solar elongations up to $20$~degrees. The PTA-class pulsars of the sample show the most significant improvements depending on the Solar elongation. For example, the rms of the residual time delays decreases of a few tens of sigma\footnote{Measured as $(\mathrm{rms}_{\mathrm{C},i}-\mathrm{rms}_{\mathrm{V},i})/\mathrm{erms}_{\mathrm{V},i}$, with $\mathrm{rms}_{\mathrm{V},i}$ and  $\mathrm{rms}_{\mathrm{C},i}$ being the rms of the residuals left by, respectively, the spherical model with time-dependent amplitude and the spherical model with time-invariable amplitude at Solar elongation $i$, and $\mathrm{erms}_{\mathrm{V},i}$ being the uncertainty to $\mathrm{rms}_{\mathrm{V},i}$.} in PSR~J0034$-$0534 below 15~degrees in Solar elongation, and from ${\sim}15$ to ${\sim}10$ sigma in PSRs~J2145$-$0750, J1022+1001, and J0030+0451 at the closest Solar elongations. 

Nevertheless, Figure~\ref{Fig:modelcomparison} also clearly shows that a simple spherical model, even when a variable amplitude is applied, does not provide a sufficiently precise description of the SW contribution to the DM. This is evident from the level of the residual time-delay's rms at the smallest elongations (lower than 10 to 20~degrees), which, in the case of the PTA-class pulsars (yielding the best DM precision), never reaches the noise floor set by the points at the largest Solar elongations. At the smallest elongations, SW acceleration leads to a steeper decrease in density than implied by a simple inverse-square. This is illustrated by \citet{bvp94} who found that a density model of $r^{-2.54}$ better fitted data inside of 10$^{\circ}$ elongation. The bimodal model proposed by \citet{yhc07a} used separate density models to account the fast and slow streams, based on results published by \citet{gf95a}, \citet{gf98} and \citet{ma81}, \citet{all47} respectively. The results presented here further demonstrate the necessity to account for SW acceleration in measurements taken close to the Sun. 

This confirms the findings of \citet{tvs19}, and extends them to a larger number of pulsars at different ecliptic latitudes.

\subsection{Temporal and latitudinal evolution of the SW density}\label{Sec:amplres}
\begin{figure*}
\begin{tabular}{cc}
\includegraphics[scale=0.41,trim={2.3cm 2cm 2cm 0}]{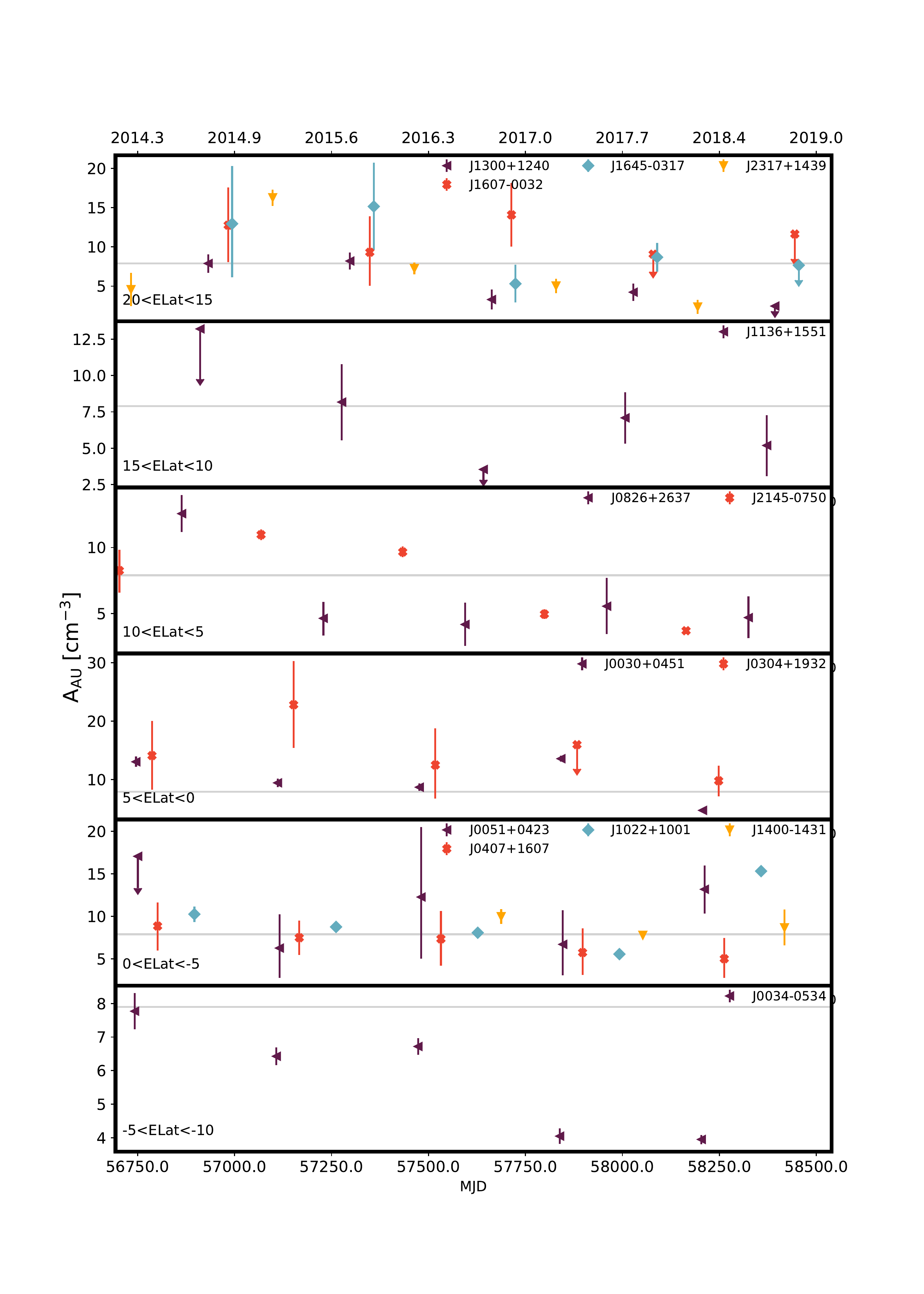} &
    \includegraphics[scale=0.41,trim={1.5cm 2cm 2cm 0}]{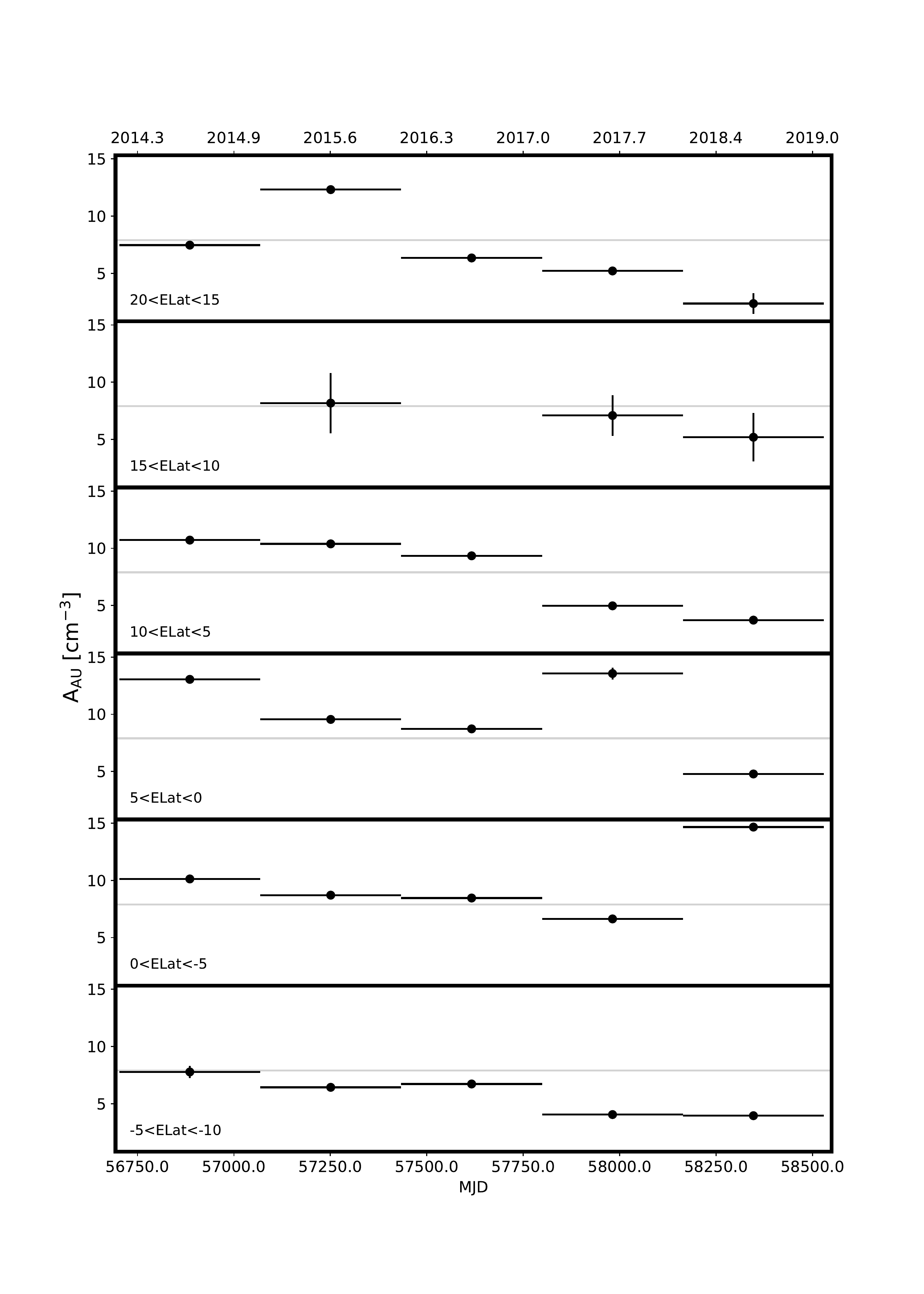}
    \end{tabular}
    \caption{Temporal trend of the amplitude for the SW spherical model, divided into bins of ecliptic latitude, for the individual pulsars (left panel) and averaged year by year (right panel). The gray line marks an amplitude of $7.9~\mathrm{cm}^{-3}$ (the best-fit value from \citealt{mca19}). All data points are shown with vertical error bars, while the upper limits are displayed at the levels of the 95th percentile of the posterior distribution in the left panel, and not considered in the right panel.}
     \label{Fig:n0timelat}
\end{figure*}

Our results show that the amplitude of the spherical SW model is not constant with time and ecliptic latitude. This is displayed in the left panel of Figure~\ref{Fig:n0timelat}, which reports the temporal evolution of the aforementioned amplitudes, each obtained from the data of a specific Solar approach for each pulsar. By averaging the computed SW amplitudes (excluding the upper limits) obtained from different pulsars in a given year and latitude range (right panel of Figure~\ref{Fig:n0timelat}), a decreasing trend in time becomes evident at ecliptic latitudes between 5 and 20 degrees, and from $-$5 to $-$10 degrees. The identified decreasing trend may be related to the cyclical Solar activity, which peaked in 2014 and is expected to have reached a minimum in 2020. Conversely, the amplitude of the model does not show an evident decrease in time for pulsars with an ecliptic latitude included between $-5$ and $+5$ degrees. This may be due to the smaller distance to the Ecliptic with respect to other pulsars, but additional studies are needed to identify the cause of these temporal trends\footnote{Because the spherical model is known to be an imperfect SW approximation \citep{tvs19}, we report in Appendix~\ref{Admtrends} an analogous overview of the DM variations.}.

\section{Future prospects for study of SW-induced DM signatures}\label{Sec:future}
Here we comment on possible future studies of this kind, specifically we identify two possible sources for future studies with LOFAR and comment on the potential for future radio observatories.
\subsection{Prospects for LOFAR Studies}
Some of the pulsar campaigns whose data we have used in this article are carried out with the LOFAR core at a monthly cadence. As stated in Section~\ref{Sec:dataset}, if a pulsar is observed at a monthly cadence only (i.e., without any coverage from the international stations, that usually observe at a weekly cadence) it has not been taken into consideration for this project. However, it is worth mentioning two particularly promising pulsars for possible SW studies, because of their low ecliptic latitude and good DM precision: PSRs J1730$-$2304 (with a median DM precision of $3\times10^{-4}~\mathrm{pc/cm^3}$ and an ecliptic latitude of 0.19~degrees) and J2256$-$1024 (with a median DM precision of $6\times10^{-5}~\mathrm{pc/cm^3}$ and an ecliptic latitude of $-3.41$~degrees). New observing campaigns with the LOFAR core have been recently carried out with the aims of increasing the observing coverage of these and other pulsars during their Solar approaches, and of collecting simultaneous data for interplanetary scintillation studies to be compared with the pulsar-based results and white-light observations, as detailed below. 

Moreover, as the decreasing trend shown in many pulsars might be related to the Solar cycle, future LOFAR observations may be able to detect an increase in the excess DM induced by the SW as the Solar activity is expected to increase in the upcoming years.

Further research is also planned to investigate more robust methods to estimate the SW contribution to pulsar DM. This includes more detailed investigation of the bimodal approach of \citet{yhc07a}, incorporating a direct comparison of the line-of-sight with Carrington maps of coronal white light to ascertain regions of fast and slow SW, as in the approach taken recently by \citet{ttt20}. Other approaches under investigation involve the use of 3-D tomographic reconstructions of velocity and density in the inner heliosphere, obtained from observations of interplanetary scintillation and coronal white light \citep[e.g][and references therein]{bbc20}, and the use of space weather models such as EUHFORIA \citep{pp17}.

\subsection{SW studies with next-generation telescopes}
While we have demonstrated the capabilities of LOFAR in monitoring the SW with pulsars, its activity could be complemented by the utilization of other telescopes. 

Observing facilities in the Southern hemisphere with the capability of covering low-frequency ranges, such as Murchinson Widefield Array \citep{tgb13} and the upcoming SKA1-Low \citep{bbg15} will have access to a different and more extended set of pulsars in comparison to LOFAR, because of a more prolonged visibility of the inner Galaxy. Furthermore, with telescopes such as CHIME \citep{baa14}, MeerKAT \citep{jon09}, SKA1-Low, LOFAR2.0 (the upcoming upgrade of LOFAR) and NenuFAR\footnote{\url{https://nenufar.obs-nancay.fr/}} \citep{bgt20} there will be access to a wider frequency bandwidth, with a consequential enabling of an increased range of scientific studies. As some of these facilities have the possibility of multi-beaming and sub-arraying, they will also be able to track more pulsars simultaneously, hence reducing the times of the observing campaigns.

\section{Conclusions}\label{Sec:conclusions}

We presented a study of the impact of the SW on a large sample of pulsars observed for up to 6 years with the LOFAR telescope. This study demonstrated that the spherically symmetric and static SW model, that is commonly used in pulsar-timing experiments, can be improved by allowing a time-variable amplitude, but that these improvements do not suffice for correcting the SW impact at the levels required for high-precision pulsar-timing experiments \citep[consistent with our earlier findings in][]{tvs19}. For PTA-class pulsars, in particular, the residuals between observations taken at small Solar elongations (lower than 10 to 20~degrees) and the corresponding model are not able to reach the noise-floor set by residuals corresponding to larger Solar elongations. As \citet{thk16} demonstrated that the SW is a potential source of false GW detections, it is then advisable to adopt time-variable SW amplitude and treat carefully observations taken during Solar conjunction within high-precision pulsar timing experiments.\\  

Moreover, the data indicated that the amplitude of the spherical model has a dependency on time for pulsars whose ecliptic latitude lies either between $-10$ and $-5$ degrees or $+5$ and $+20$ degrees. In particular, the amplitude decreases from the first half of the observed time-span to the second half of that time-span. On the other hand, the amplitude tends to remain constant for pulsars with an ecliptic latitude included between $-5$ and $+5$ degrees. Additional studies will be necessary to recognize the cause of these trends, and whether they are in connection to the Solar cycle activity.

While the investigation of new SW models for high-precision pulsar timing was not in the scope of the current work, a series of measures can be adopted to mitigate the SW-induced noise (assuming the absence of frequency-dependent DM or scattering):

\begin{itemize}
    \item Carrying out observations with wideband observing receivers may allow, in combination with the usage of frequency-resolved templates, a precise determination of time-dependent DM (due to the IISM, the SW or both) that can be then used to correct the noise induced by variable dispersion;
    \item Carrying out simultaneous observations at low and high frequency to allow a precise DM determination by using the low-frequency data. This can be then used to correct the high-frequency ToAs;
    \item If obtaining simultaneous observations is not possible, it may be of aid carrying out high-cadence, low-frequency observations (ideally once every two or three days) for two weeks around the Solar conjunction. The DM values obtained from these observations can be interpolated and used as correction scheme for high-frequency observations. The high-cadence of the observations is important in this case because, as shown by \citet{nhw17}, the fast SW variability may invalidate the DM corrections if the high- and low-frequency observations are separated by more than one day;
    \item For pulsars with a flat spectral index, it may be meaningful to carry out observations at very high frequencies (at S band or more), where the DM-induced noise is marginal, and apply the spherical SW model by choosing an amplitude accordingly to, e.g., the results of this article depending on the pulsar's ecliptic latitude. However, this approach implies that a certain amount of timing noise will be left in the data, especially the ones taken close to the Solar conjunction. Moreover, \citet{lmc18} demonstrated that observing frequencies lower than 1~GHz are more optimal to achieve high precision in pulsar timing experiments.
    \end{itemize}

\section*{Data Availability}

The data underlying this article that were collected with the International LOFAR Stations and with the LOFAR core under still private observing programs, will be shared on reasonable request to the corresponding author. The data underlying this article collected with the LOFAR core and under public observing programs are available at: \url{https://lta.lofar.eu/}.\\
The initial timing models, the ToA files and the DM time series for the pulsars used in this article are publicly available on Zenodo from the 1st of January 2021 (DOI: 10.5281/zenodo.4247554).

\begin{acknowledgements}
This work is part of the research program Soltrack with project number 016.Veni.192.086, which is partly financed by the Dutch Research Council (NWO).

The authors thank the anonymous referee for their support and useful comments. CT 

This paper is partially based on data obtained with: i) the German stations of the International LOFAR Telescope (ILT), constructed by ASTRON \citep{vwg13} and operated by the German LOng Wavelength (GLOW) consortium (\url{https://www.glowconsortium.de/}) during station-owners time and proposals LC0\_014, LC1\_048, LC2\_011, LC3\_029, LC4\_025, LT5\_001, LC9\_039, LT10\_014; ii) the LOFAR core, during proposals LC0\_011, DDT0003, LC1\_027, LC1\_042, LC2\_010, LT3\_001, LC4\_004, LT5\_003, LC9\_041, LT10\_004, LPR12\_010; iii) the Swedish station of the ILT during observing proposals carried out from May 2015 to January 2018. We made use of data from the Effelsberg (DE601) LOFAR station funded by the Max-Planck-Gesellschaft; the Unterweilenbach (DE602) LOFAR station funded by the Max-Planck-Institut f\"ur Astrophysik, Garching; the Tautenburg (DE603) LOFAR station funded by the State of Thuringia, supported by the European Union (EFRE) and the Federal Ministry of Education and Research (BMBF) Verbundforschung project D-LOFAR I (grant 05A08ST1); the Potsdam (DE604) LOFAR station funded by the Leibniz-Institut f\"ur Astrophysik, Potsdam; the J\"ulich (DE605) LOFAR station supported by the BMBF Verbundforschung project D-LOFAR I (grant 05A08LJ1); and the Norderstedt (DE609) LOFAR station funded by the BMBF Verbundforschung project D-LOFAR II (grant 05A11LJ1). The observations of the German LOFAR stations were carried out in the stand-alone GLOW mode, which is technically operated and supported by the Max-Planck-Institut f\"ur Radioastronomie, the Forschungszentrum J\"ulich and Bielefeld University. We acknowledge support and operation of the GLOW network, computing and storage facilities by the FZ-J\"ulich, the MPIfR and Bielefeld University and financial support from BMBF D-LOFAR III (grant 05A14PBA) and D-LOFAR IV (grants 05A17PBA and 05A17PC1), and by the states of Nordrhein-Westfalia and Hamburg. MB acknowledges support from the Deutsche Forschungsgemeinschaft under Germany's Excellence Strategy - EXC 2121 "Quantum Universe" - 390833306. CT acknowledges support from Onsala Space Observatory for the provisioning of its facilities/observational support. The Onsala Space Observatory national research infrastructure is funded through Swedish Research Council grant No 2017-00648.

GS was supported by the Netherlands Organization for Scientific Research NWO(TOP2.614.001.602). JPWV acknowledges support by the Deutsche Forschungsgemeinschaft (DFG)
through the Heisenberg program (Project No. 433075039).
\end{acknowledgements}

\begin{appendix}
\section{Effectiveness of the IISM-SW disentangling scheme}\label{Asimulations}
\begin{figure}
    \centering
    \includegraphics[scale=0.34,trim={2cm 0cm 0cm 0cm}]{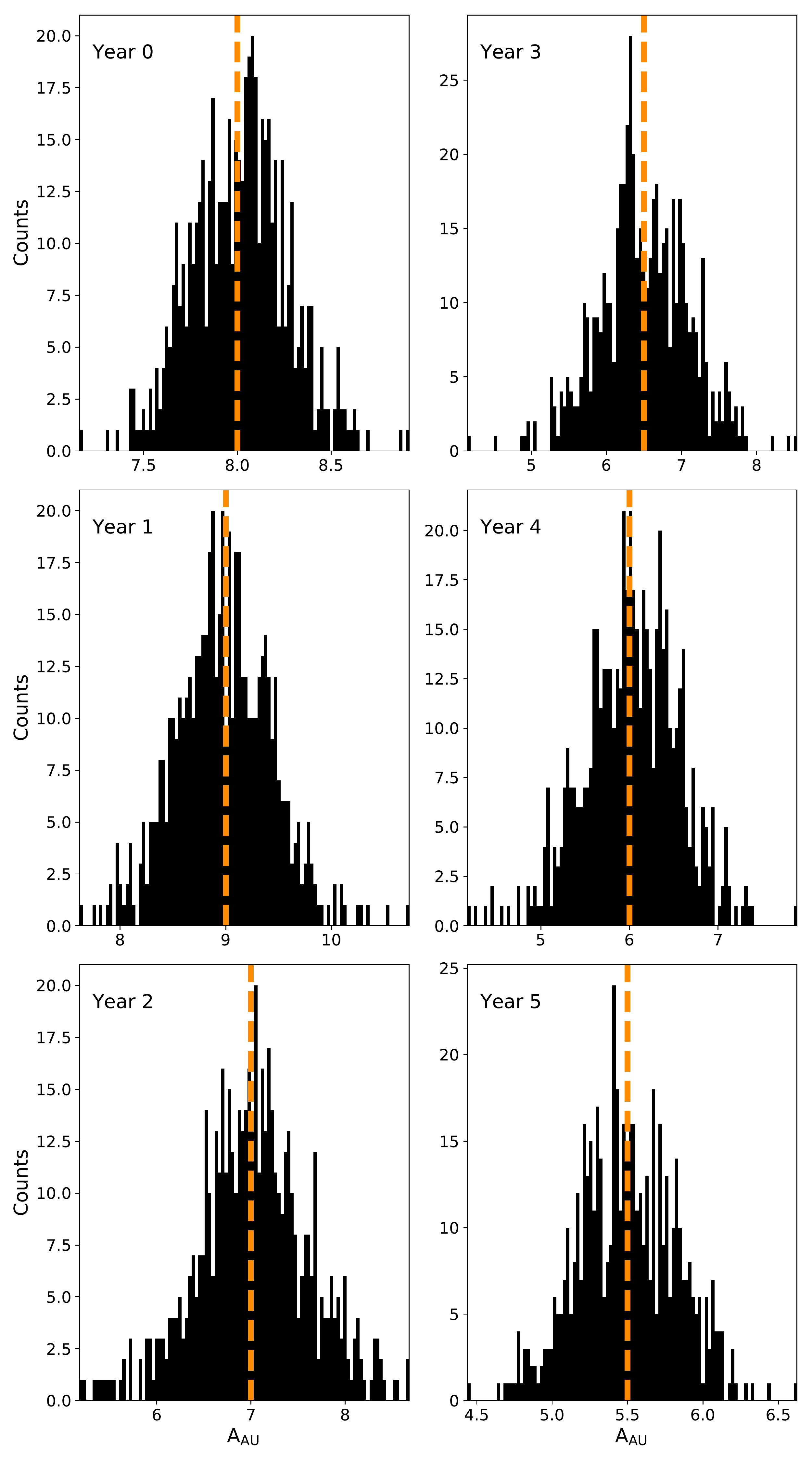}
    \caption{Injected (dashed, orange lines) and recovered (filled, black histograms) values of the amplitude of the spherically symmetric SW model.}
    \label{Fig:polyIISMsimulation}
\end{figure}

To confirm that our choice of a third order polynomial is a sufficient description of the IISM contributions, we have simulated 500 DM time series affected by Kolmogorov turbulence and white noise as drawn from a zero mean Gaussian population with a variance of $1\times 10^{-4}~$ pc/cm$^3$, and we have attributed irregular error bars to match the ones of PSR~J1022+1001. To this time-series, we then added the SW contribution from a spherically symmetric model, with a variable amplitude per year. This final dataset was then analyzed following the steps detailed in Section~\ref{Sec:disentangl}. For each year and injected value of the spherical SW amplitude, as shown in Figure~\ref{Fig:polyIISMsimulation}, we were able to recover a statistically comparable value of the amplitude of spherical SW model. 

We further tested whether the third order polynomial model is able to account for the IISM contribution only. To this end, we repeated the above simulations without adding the SW contribution, and we analyzed the DM time series using the MCMC algorithm described in Section~\ref{Sec:disentangl}, modified to apply a segment-by-segment cubic-only model. To determine the success rate of modeling procedure we used a two-tailed Kolmogorov-Smirnov (KS) test. The results report that in the 100\% of the cases, the distribution of the residuals after applying the cubic model are identical to a zero-mean Gaussian distribution. This confirms the ability of our algorithm in removing the IISM contribution from the DM time series.

\begin{figure*}
\centering
    \begin{tabular}{cc}
    \includegraphics[scale=0.34,trim={0cm 0cm 0cm 0cm}]{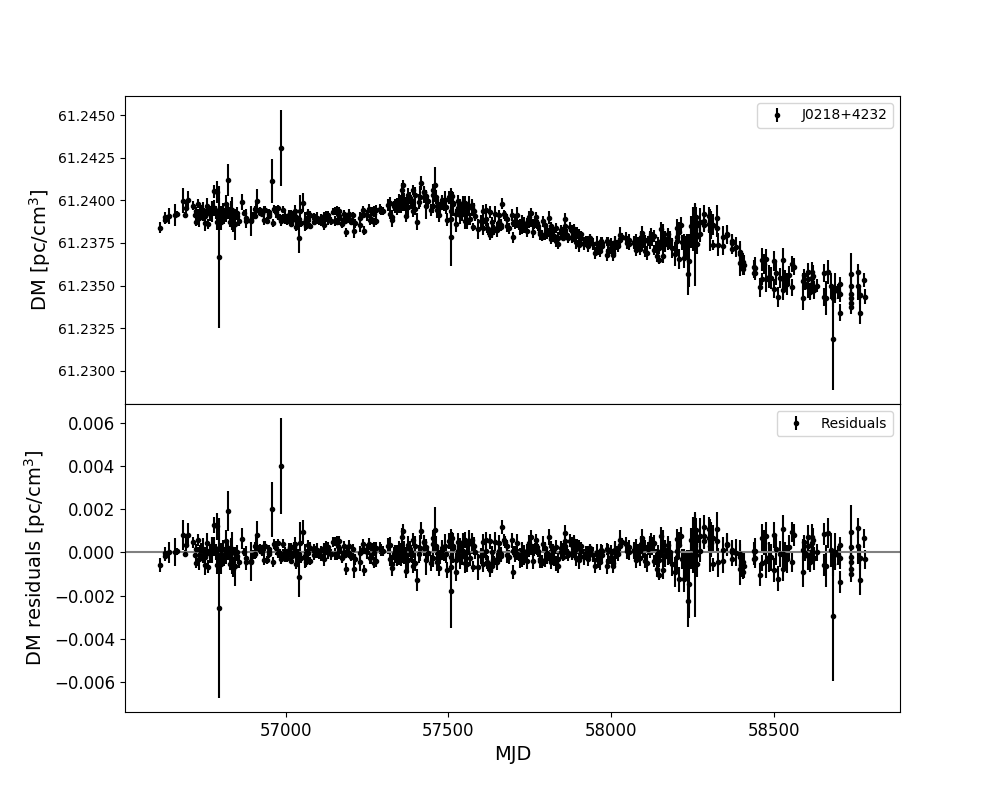} & \includegraphics[scale=0.34,trim={0cm 0cm 0cm 0cm}]{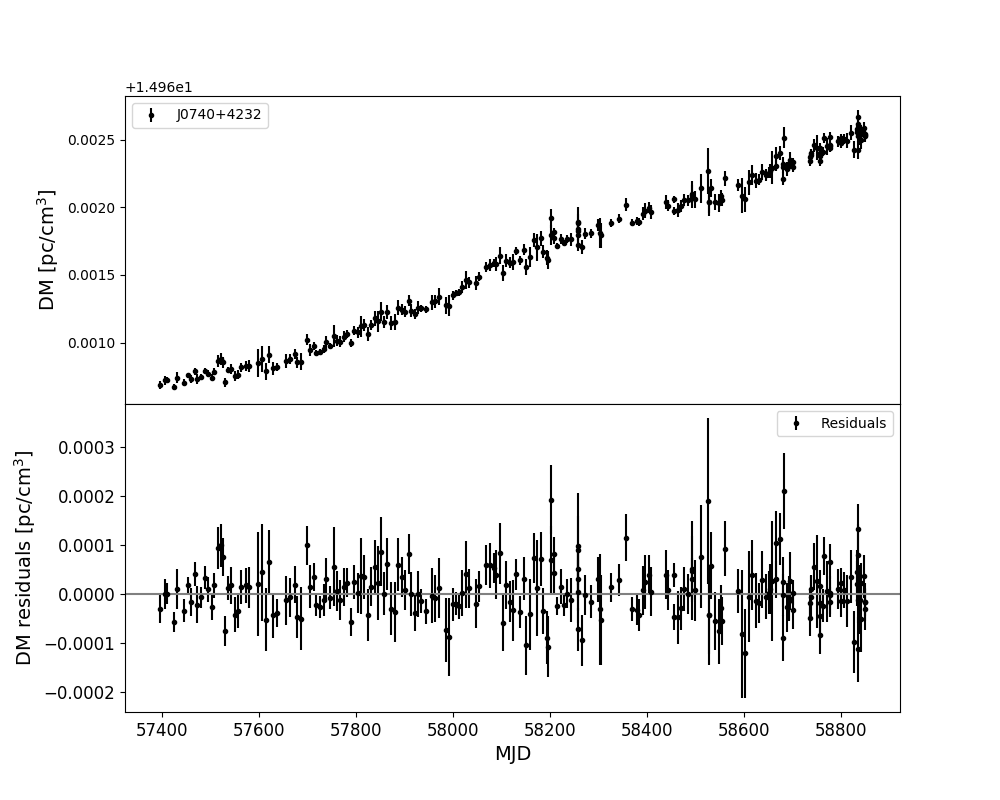}\\
      \includegraphics[scale=0.34,trim={0cm 0cm 0cm 0cm}]{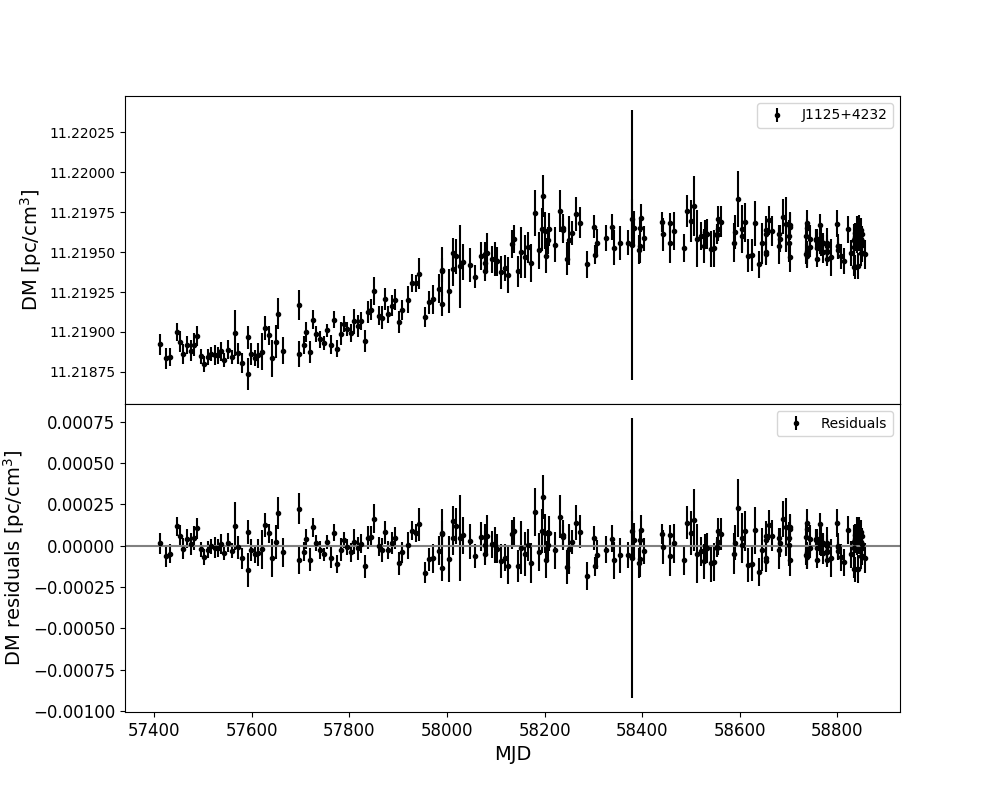} & \includegraphics[scale=0.34,trim={0cm 0cm 0cm 0cm}]{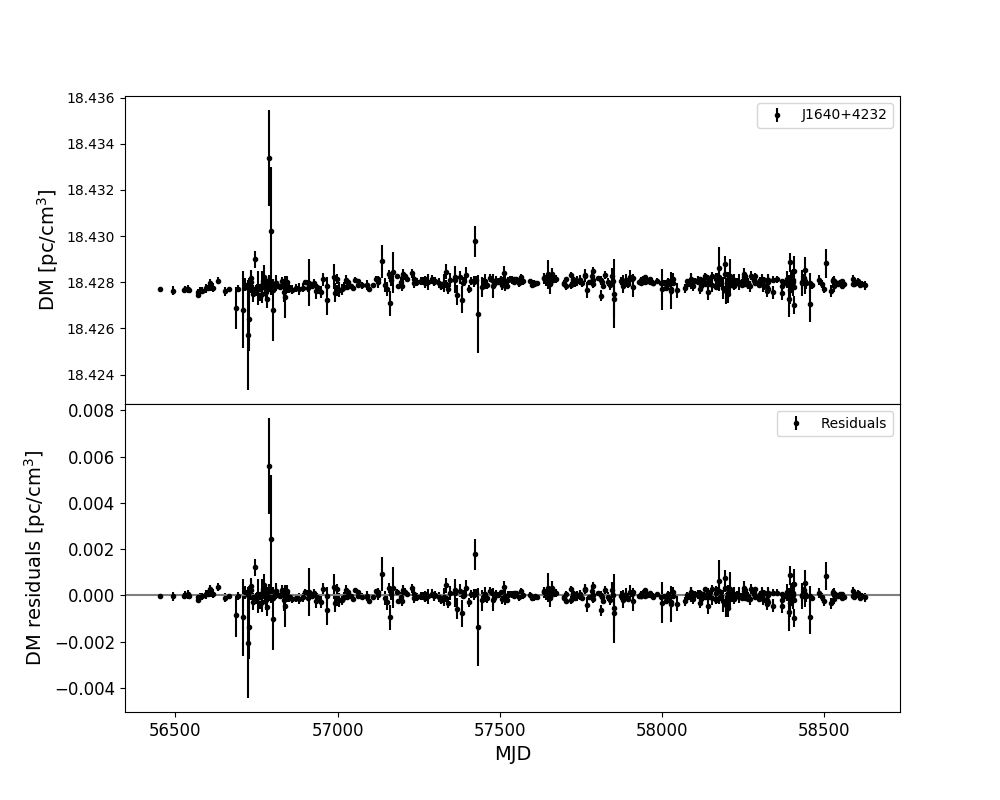}\\
    \end{tabular}
    \caption{DM time series of the four tested millisecond pulsars (upper panel of each quadrant) and their post-modeling residuals (bottom panels).}
    \label{Fig:4msps}
\end{figure*}

To test whether our algorithm is as performing on real, irregular DM time series as it is on simulated data, we repeated the just-described analysis on the DM time series of 4 millisecond pulsars presented in \citet{dvt20}, namely PSRs~J0218+4232, J0740+6620, J1125+7819, J1640+2224. These sources have a high ecliptic latitude ($>30$ degrees); therefore, we can assume that they are only affected by IISM-induced variations. We used the IISM-only MCMC algorithm that we described in the previous paragraph (i.e., modified to apply a segment-by-segment cubic-only model) to model the IISM variations presented by these pulsars. The two-tailed KS test previously described showed that, after modeling the DM time series with the results of our algorithm, the residuals resulted as well compatible with a zero-mean Gaussian distribution (see Figure~\ref{Fig:4msps}).

Thus we are able to demonstrate that our modeling method leads to the IISM contribution being successfully disentangled from the SW component.

\section{Pulsars without significant SW signatures}\label{Anoswsignature}
Table~\ref{Tab:sources_nosw} reports the investigated pulsars that do not show significant SW signatures.
\begin{table*}
\centering
\caption{Investigated pulsars that do not show significant SW signatures. The table reports the source name, the covered time-span, the number of observing sites that have been monitoring that specific source, the Galactic coordinates, the rotational spin period, ecliptic latitude, dispersion measure (DM) of the pulsar as measured during the general pulsar timing analysis described in Section~\ref{Sec:dataanalysis}, the decimal logarithm of the median DM uncertainty and the number of observations used to generate the data-derived template.}\label{Tab:sources_nosw}
{\small
\begin{tabular}{cccS[table-format=2.1]S[table-format=2.1]S[table-format=4.1]S[table-format=2.2]S[table-format=2.1]S[table-format=2.2]S[table-format=3]}
\hline
Name & Time-span & Observing & \multicolumn{2}{c}{Galactic} &\multicolumn{1}{c}{Period} & \multicolumn{1}{c}{Ecliptic} & \multicolumn{1}{c}{DM} & \multicolumn{1}{c}{Log} & \multicolumn{1}{c}{Observations}\\
&           & Sites & \multicolumn{2}{c}{Coo. [deg]} &\multicolumn{1}{c}{[ms]} & \multicolumn{1}{c}{Latitude [deg]} & \multicolumn{1}{c}{[pc/cm$^3$]} & \multicolumn{1}{c}{M(eDM)} & \multicolumn{1}{c}{per template} \\
\hline
J0034$-$0721 &	2013-08	\quad 2019-08&   7& 110.4&-69.8	&	943.0 & -10.15	& 10.9   & -2.57  & 215         \\ 
J0137+1654 		  &	2013-09	\quad 2019-05&   6&	 138.4&-44.6	&	414.8 & 6.31 	& 26.1   & -2.64  & 199        \\ 
J0151$-$0635 &	2013-09	\quad 2019-05&   5&	 160.4&-65.0	&	1464.7 & -16.84 & 25.8   & -2.02  & 234          \\ 
J0525+1115  &	2013-09	\quad 2019-05&   5&	 192.7&-13.2	&	354.4 & -11.93 	& 79.4   & -2.52  & 207         \\ 
J0528+2200	  &	2013-08	\quad 2019-08&   6& 183.9&-6.9	&	3745.5 & -1.24 	& 50.9   & -2.84  & 181         \\ 
J0538+2817 		  &	2014-02	\quad 2019-04&   6&	 179.7&-1.7	&	143.2 & 4.94 	& 40.0   & -2.15  & 190        \\ 
J0540+3207 		  &	2016-03	\quad 2019-05&   5&	 176.7&0.8	&	524.3 & 8.76 	& 62.1   & -3.13  & 122        \\ 
J0543+2329	  &	2013-08	\quad 2019-04&   5&	 184.4&-3.3	&	246.0 & 0.10 	& 77.7   & -2.39  & 237        \\ 
J0609+2130 		  &	2013-10	\quad 2019-05&   7& 189.2&1.0	&	55.7 & -1.92 	& 38.7   & -2.91  & 197        \\ 
J0612+3721	  &	2013-09	\quad 2019-05&   6&	 175.4&9.1	&	298.0 & 13.95 	& 27.2   & -2.99  & 216        \\ 
J0614+2229	  &	2013-08	\quad 2019-08&   5& 188.8&2.4	&	335.0 & -0.90 	& 96.9   & -2.98  & 196        \\ 
J0629+2415 	  &	2013-08	\quad 2019-05&   4&	 188.8&	6.2	&	476.6 & 0.99 	& 84.2   & -2.98  & 189        \\ 
J0659+1414	  &	2013-09	\quad 2019-08&   5& 201.1&	8.3	&	384.9 & -8.44 	& 14.1   & -1.96  & 251        \\ 
J0823+0159	  &	2013-08	\quad 2019-05&   4&	 222.0&	21.2	&	864.9 & -16.90 	& 23.8   & -2.78  & 220         \\ 
J0837+0610 	  &	2013-08	\quad 2019-05&   6&	 219.7&	26.3	&	1273.8 & -11.98 & 12.9   & -3.88  & 202          \\ 
J0943+1631 	  &	2013-08	\quad 2019-05&   5&	 216.6&	45.4	&	1087.4 & 2.70 	& 20.3   & -1.77  & 255        \\ 
J0953+0755 	  &	2013-08	\quad 2019-05&   7 & 228.9&	43.7	&	253.1 & -4.62 	& 3.0    & -3.27  & 276        \\
J1024$-$0719 		  &	2012-12	\quad 2019-08&   1& 251.7&	40.5	&	5.2 & -16.04 	& 6.5    & -3.32  & 68      \\
J1543$-$0620 &	2013-12	\quad 2019-05&   6& 0.6 &	36.6	&	709.1 & 13.06 	& 18.4   & -3.49  & 216        \\
J1705$-$1906 &	2016-03	\quad 2019-04&   4&	 3.2 &	13.0	&	299.0 & 3.72 	& 22.9   & -2.3   & 153      \\
J1744$-$1134 		&	2012-12	\quad 2019-08 & 7 & 14.8& 9.2    & 4.1 	& 11.80 & 3.1   & -3.77 &394  \\
J1820$-$0427 &	2013-08	\quad 2019-05&   4&	 25.5&	4.7	&	598.1 & 18.88 	& 84.3   & -2.54  & 211       \\
J1825$-$0935 &	2013-12	\quad 2019-08&   5& 21.4&	1.3	&	769.0 & 13.71 	& 19.4   & -3.29  & 184        \\
J1834$-$0426 &	2013-08	\quad 2019-05&   5&	 27.0&	1.7	&	290.1 & 18.73 	& 79.4   & -2.63  & 156       \\
J1848$-$1952 &	2016-03	\quad 2019-05&   5&	 14.8&	-8.3	&	4308.2 & 3.09 	& 18.3   & -2.25  & 103        \\
J1900$-$2600 &	2013-08	\quad 2019-05&   5&	 10.3&	-13.5	&	612.2 & -3.29 	& 37.9   & -2.58  & 218       \\
J1913$-$0440 &	2013-09	\quad 2019-08&   6& 31.3&	-7.1	&	825.9 & 17.53 	& 89.4   & -2.83  & 1          \\
J2051$-$0827 		  &	2013-06	\quad 2019-03&   2& 39.2&	-30.4	&	4.5 & 8.85 	& 20.7   & -3.58  & 67      \\
J2222$-$0137 		  &	2016-03	\quad 2019-05&   6& 62.0&	-46.1	&	32.8 & 7.98 	& 3.3    & -3.56  & 37      \\

\hline 
\end{tabular}
}
\end{table*}

\section{Variations in the SW impact on the DM time series}\label{Admtrends}
\begin{figure*}
    \centering
    \includegraphics[scale=0.48,trim={3cm 2cm 2cm 2cm}, angle=-90 ]{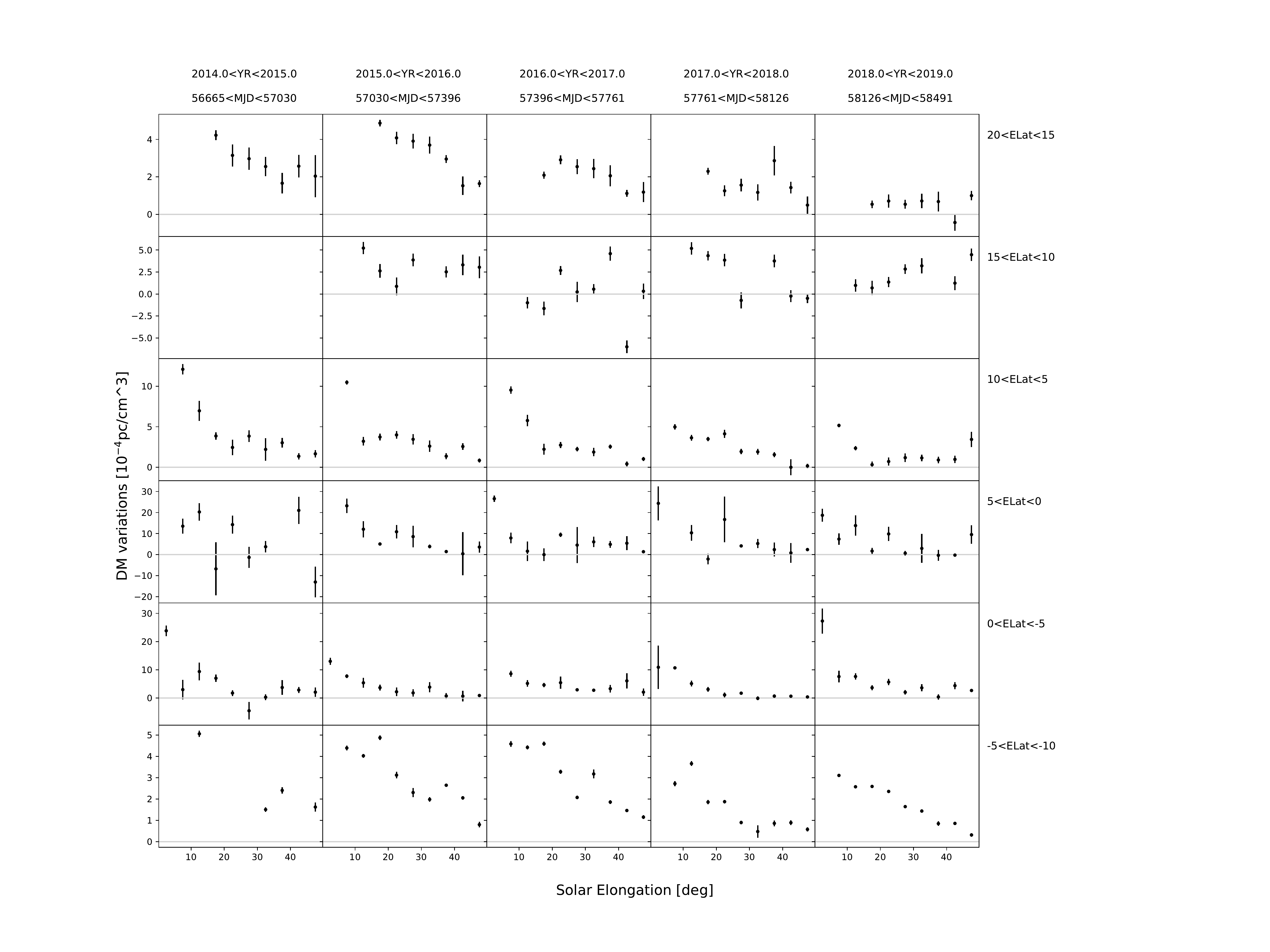}
    \caption{Average DM variations (after the subtraction of the IISM) as a function of Solar elongation, divided per time and ecliptic latitude. All data points are shown with vertical error bars.}
    \label{Fig:dmtimeelat}
\end{figure*}

Section~\ref{Sec:amplres} reports the amplitude of the spherical SW model as variable across the years for a number of pulsars. However, \citet{tvs19} (and this article) showed the spherical model to be an imperfect SW approximation. As an additional confirmation of the detected variability of the SW contribution, we also examined the DM variations themselves, after subtracting the IISM component. Figure~\ref{Fig:dmtimeelat} shows the averages of DM variations as a function of the Solar elongation (in 5 degree bins, up to 50 degrees) from all the 14 pulsars, organized by ecliptic latitude and time of observation. As for the amplitudes of the spherical SW model, it is possible to verify that also the magnitude of the DM variations decreases in time within the ecliptic latitude ranges spanning from 20 to 5 degrees and $-$5 to $-$10 degrees. For example, for pulsars with ecliptic latitude ranging from 15 and 20 degrees, the average DM variations at Solar elongations smaller than 20 degrees decrease from $4.2\times10^{-4}$ pc/cm$^{-3}$ to $5\times10^{-5}$ pc/cm$^{-3}$. Similarly, pulsars whose ecliptic latitudes range between 5 and 10 degrees have DM variations at Solar elongations lower than 10 degrees that drop from $1.2\times10^{-3}$ pc/cm$^{-3}$ to $5.2\times10^{-4}$ pc/cm$^{-3}$ across the timespan. On the other hand, the DM variations for pulsars with ecliptic latitudes included between $-$5 and 5 degrees tend to fluctuate without showing a clear decreasing trend. This confirms the trend seen in Figure~\ref{Fig:n0timelat}.

\end{appendix}

%
   \bibliographystyle{aa} 
  \bibliography{psrrefs,journals,modrefs,crossrefs} 
%

\end{document}